\documentstyle[12pt,amssymb]{article}
\begin{document}

\markright{Immirzi Ambiguity for Black Holes}

\begin{center}
{\large \bf Immirzi Ambiguity, Boosts and Conformal Frames for
Black Holes}
\\[3mm] {Luis J Garay and Guillermo A Mena Marug\'{a}n}
\\[3mm] {\it  Centro de F\'{\i}sica Miguel A Catal\'{a}n, IMAFF, CSIC,
\\ Serrano 121, 28006 Madrid, Spain}
\end{center}

\begin{abstract}
We analyze changes of the Immirzi parameter in loop quantum
gravity and compare their consequences with those of Lorentz
boosts and constant conformal transformations in black hole
physics. We show that the effective value deduced for the Planck
length in local measurements of vacuum black holes by an
asymptotic observer may depend on its conformal or Lorentz frame.
This introduces an apparent ambiguity in the expression of the
black hole entropy which is analogous to that produced by the
Immirzi parameter. For quantities involving a notion of energy,
the similarity between the implications of the Immirzi ambiguity
and a conformal scaling disappears, but the parallelism with
boosts is maintained. \vskip 3mm \noindent {PACS numbers:04.70.Dy,
04.60.Ds, 03.30.+p}
\end{abstract}

Almost thirty years ago, a combination of classical and
semiclassical arguments led to the conclusion that black holes
possess thermodynamic properties \cite{Wald}-\cite{Thaw}. In
particular, a study of the mass change produced by infinitesimal
variations of the black hole parameters revealed a close
similarity with the first law of thermodynamics if one identifies
the entropy $S$ and temperature $T$ with the area $A$ and surface
gravity $\kappa$ of the black hole horizon \cite{BCH}. Based
partly on these results, Bekenstein proposed to regard $A$ as
really providing a gravitational entropy \cite{Tbec}. At that
time, however, no radiative process associated with a black hole
was known from which one could infer a physical notion of
temperature. Besides, the formula for the mass variation
\cite{BCH} allowed to assign the formal roles of temperature and
entropy only up to an ambiguity in a constant \cite{Wald}. In this
context, Hawking performed a semiclassical analysis using methods
of quantum field theory in curved backgrounds \cite{Thaw}. He
proved that the presence of a black hole results in a thermal
radiation with $T=\kappa \hbar/(2\pi c)$. Here, $\hbar$ and $c$
are the Planck constant and the speed of light (and the Boltzmann
constant has been absorbed in the definition of $T$). Hence, a
black hole has an entropy determined by the Bekenstein-Hawking
(BH) formula $S=A/(4l_P^2)$, where $l_P=\sqrt{\hbar G/c^3}$ is the
Planck length and $G$ is the Newton constant.

A statistical explanation for this entropy has been only recently
suggested in terms of the number of quantum states corresponding
to a macroscopic black hole configuration
\cite{BraCo,Eash}. Among the explanations proposed, we will
concentrate our attention on that inspired in loop quantum gravity
\cite{Eash}. In this case, the states counted are spin networks
compatible with the horizon geometry. It is generally admitted
that this derivation of the entropy formula is affected by an
ambiguity in a positive factor. In fact, loop gravity is based on
the Ashtekar formulation of general relativity \cite{Ash} in which
the phase space is described by a real $SU(2)$ connection and a
densitized triad. This real connection can be obtained as the
difference between the spin connection of the triad and the
extrinsic curvature (in triadic form) \cite{Ash}. Nonetheless, a
completely equivalent classical description is reached if, in the
above construction, one multiplies the extrinsic curvature by a
positive constant $\beta$ and divides the densitized triad by the
same factor. This factor is usually called the Immirzi parameter
\cite{Immi}.

Regardless of the value of $\beta$, one can quantize the theory
representing quantum states as spin networks \cite{SN}. In this
framework, one can promote the area of a surface to a positive
operator that has a discrete spectrum and is well defined at least
at the kinematical level \cite{RS,Aspec}. For black holes, the
boundary conditions ensure that the operator for the horizon area
is also dynamically consistent, providing an observable
\cite{Eash,Cash}. The remarkable fact realized by Immirzi is that
quantization procedures based on connections that differ in the
value of $\beta$ lead to different area spectra \cite{Immi}.
Actually, the Immirzi parameter appears as a global factor in the
eigenvalues of the horizon area, and it is only the spectrum of
$A_{\beta}/\beta$ that is independent of the selected
quantization. We have introduced the notation $A_{\beta}$ to
indicate that the area is calculated in the formulation with
Immirzi parameter equal to $\beta$. The counting of quantum states
then leads to the black hole entropy $S=\beta_0 A_{\beta}/(4\beta
l_{\star}^2)$ \cite{Eash}, where $\beta_0=\ln{2}/(\pi\sqrt{3})$
and the length $l_{\star}$ is determined by the fact that the
Hilbert-Einstein (HE) term in the gravitational action is divided
by $16\pi l_{\star}^2/(c\hbar)$. Given this definition,
$l_{\star}$ is commonly identified with $l_P$. In any case, it is
clear that one must demand $\beta l_{\star}^2=\beta_0 l_P^2$ to
recover the BH formula.

The roots and significance of the Immirzi ambiguity are not fully
understood \cite{Imvar}. As far as the kinematics of general
relativity is concerned, it is equivalent to the ambiguity in a
constant scale transformation, which can be viewed as a change of
conformal frame with constant conformal factor \cite{LG}.
Nevertheless, this equivalence is broken when the Hamiltonian
dynamics is taken into account. In addition, the Immirzi ambiguity
has been argued to arise from the partial gauge fixing performed
in Ashtekar gravity in terms of real connections, where a time
gauge is chosen, reducing the gauge group from $SO(3,1)$ to
$SU(2)$ \cite{Alex}. One of the reasons for the confusion about
the origin of this ambiguity is that, while the black hole entropy
depends on the Immirzi parameter in loop quantum gravity, no
ambiguity seems to arise in the standard deduction of the BH
formula \cite{BD}. Our aim in this work is to show that, contrary
to this belief, there exist in fact certain ambiguities which,
rather than the value of the entropy, affect the empirical
definition of the Planck length. As a result, this length becomes
in practice an effective scale whose value (in terms of
$l_{\star}$) may vary with the frame employed. We will consider
frames that are related by constant conformal transformations and
inertial frames in the asymptotic flat region of the black hole
which are not at rest. We will analyze the role of these conformal
transformations and boosts in black hole physics and compare them
with a change of the Immirzi parameter in Ashtekar gravity.

Let us first study the effect of conformal transformations. We
consider two frames with line elements related by
$d\tilde{s}^2=\Omega^2 ds^2$, where the conformal factor
$\Omega>0$ is constant. Obviously, this factor can be absorbed by
a dilatation of the unit of length, because the line element has
dimension of area. Hence, we can think of $ds^2$ and
$d\tilde{s}^2$ as representing the same spacetime but in systems
of coordinates that simply differ in their length scale. One would
thus expect that the predictions and measurements of two
observers, each regarding his respective conformal frame as the
physical one, were related just as dimensional analysis suggests.
Let us show that this expectation holds for black holes.

It is trivial but lengthy to repeat the semiclassical derivation
of the BH formula in the new conformal frame $d\tilde{s}^2$.
Nonetheless, we can easily understand the main consequences of
this change. Suppose $ds^2$ is the line element of a black hole
spacetime with unit timelike Killing field $\xi^{\mu}$ with
respect to which we define the black hole mass $M$. In the new
frame, the unit Killing field is
$\tilde{\xi}^{\mu}=\xi^{\mu}/\Omega$ and, as a consequence of this
new normalization, the mass transforms as
$\widetilde{\!M}=M/\Omega$, i.e., exactly as a scaling of the unit
of length would dictate (keeping $\hbar$ and $c$ fixed). The
Hawking temperature $T$, on the other hand, can be obtained, up to
a constant factor, as the inverse of the period in imaginary time
of the scalar Green functions for the black hole background
\cite{BD}. A conformal transformation of coordinates, and in
particular of time, makes this period scale with $\Omega$.
Therefore, $\widetilde{T}=T/\Omega$. Note that the black hole mass
and temperature transform in the same way, as expected since they
both have dimensions of energy. The differential mass formula
$dM=T\,dS$ (here restricted to the neutral non-rotating case for
simplicity) and its counterpart for the new conformal frame then
imply that the black hole entropy is invariant, $\widetilde{S}=S$,
confirming what one would anticipate on dimensional grounds.

Obviously, the horizon area scales as the line element under a
constant conformal transformation, $\widetilde{A}=\Omega^2 A$.
Taking into account the invariance of the black hole entropy
$\widetilde{S}=S=A/(4l_P^2)$, we see that the effective value
deduced for the Planck length in the new conformal frame is
$\widetilde{l}_P=\Omega l_P$, just what one could expect from a
dilatation. Assuming that $\hbar$ and $c$ are fundamental
constants, in the sense that they are frame-independent, we can
view this transformation as a change in the effective dimensionful
Newton constant. Actually, suppose that we define the
gravitational action adopting a certain conformal frame as a
universal frame of reference, e.g. $ds^2$. The coefficient of the
HE term is $D\equiv \hbar c/(16\pi l_{\star}^2)$, and the value of
the Newton constant is $G=c^4/(16\pi D)=l_{\star}^2c^3/\hbar$ in
this frame. Under a constant scaling $ds^2=d\tilde{s}^2/\Omega^2$,
the HE term gets divided by $\Omega^{2}$. The effective
dimensionful Newton constant in the new conformal system is then
$\widetilde{G}=\Omega^2 G$, and hence $\widetilde{l}_P=\Omega
l_P$.

This result may seem striking because it is frequently argued that
the value of the Planck length is experimentally unambiguous
(although there is an open debate about the significance of
dimensionful constants in physics \cite{dcon}). Let us analyze the
determination of $l_P$ to clarify this issue. Accepting that $c$
and $\hbar$ are fundamental constants, we can introduce a system
of units by only adopting a time unit (e.g. the second $sec$),
because its product by $c$, and $\hbar/c^2$ times its inverse,
provide length and mass units. We may define the time unit as a
multiple of the period of an atomic transition (e.g. between two
hyperfine levels of the ground state of $^{133}{\rm Cs}$). Then,
we can measure the experimental value of Newton constant, which in
our system of units will be given by a dimensionless number
multiplied by $(sec)^2c^5/\hbar$. If we change to another
conformal frame, as we have discussed above, periods will get
scaled by the conformal factor $\Omega$ and, as a consequence, the
time unit defined by the new observer will differ from the
original one by this factor ($\widetilde{sec}=\Omega\,sec$). The
measured dimensionful Newton constant (not its dimensionless
value) will then be scaled by $\Omega^2$, and the deduced Planck
length will indeed be $\widetilde{l}_P=\Omega l_P$.

In conclusion, although a change of conformal frame that amounts
to a constant scaling leaves invariant the value of the black hole
entropy, it introduces a certain degree of ambiguity in the
expression of this entropy as a function of the horizon area,
namely $S=\tilde{A}/(4\Omega^2 l_P^2)$, where $l_P$ is the Planck
length in the chosen frame of reference $ds^2$ and $\tilde{A}$ is
the horizon area measured in the frame $d\tilde{s}^2=\Omega^2
ds^2$.

We now discuss the effect of boosts, for which we consider two
observers $O$ and $O^{\prime}$ in the asymptotic flat region of a
black hole spacetime. Both observers employ systems of
asymptotically Cartesian coordinates where they are at rest,
$(t,\vec{x})$ and $(t^{\prime},\vec{x}^{\,\prime})$, with
$\partial_t$ being a Killing field. The observer $O^{\prime}$
moves with constant velocity $\vec{V}$ with respect to $O$, so
that the systems are related by a boost with parameter
$\gamma=1/\sqrt{1-(V/c)^2}$ \cite{SR}, where $V=|\vec{V}|$ is the
Cartesian norm of $\vec V$. In general, this Lorentz
transformation leaves the spacetime metric invariant only in the
asymptotic flat region. To study the consequences of Lorentz
invariance, we will thus restrict our analysis to that region.
Furthermore, we will disregard measurements that require an
infinite amount of time for the transmission of the collected
information to the observer (through the asymptotic flat region),
because they are not feasible \cite{Note}. In practice, this means
that we will only consider local measurements. This rationale
eliminates, for instance, observations of the black hole effects
made from different angles in the asymptotic region, because such
measurements would imply infinite spatial separations.

Local measurements in the asymptotic flat region that indicate the
presence of a vacuum black hole are those that detect the outward
flux of Hawking radiation. The two observers $O$ and $O^{\prime}$
will see a thermal flux of particles, but their records will
differ in the apparent temperature. The difference is due to the
Doppler redshift that frequencies experiment in the boosted frame
with respect to $O$ \cite{BD}, i.e., with respect to the observer
whose time coordinate corresponds to a Killing field. It is known
that frequencies are shifted by a boost according to the law
$\nu^{\prime}=\alpha\nu$ \cite{SR}, where
\begin{equation}\label{redshift}
\alpha=\gamma(1-c^{-1}\,V\cos{\varphi})\end{equation} and
$\varphi$ is the angle formed by $\vec{V}$ and the direction of
propagation in the rest frame of the source. In our case, Hawking
radiation propagates radially and its rest frame corresponds to
that of $O$. Note that, from our comments above, $\varphi$ is a
constant in the region where the observer $O^{\prime}$ performs
her (local) measurements. On the other hand, if we let $\vec{V}$
vary by analyzing different boosted observers, $\alpha$ may take
any positive real value.

Owing to this redshift, the radiation detected by $O^{\prime}$
possesses the apparent temperature $T^{\prime}=\alpha T$. The same
redshift affects also the energy density observed in the presence
of the Hawking flux, $\rho^{\prime}=\alpha\rho$. To show this
fact, we consider the stress-tensor in the Unruh vacuum that
describes the Hawking evaporation process, $\langle
T_{\mu\nu}\rangle$ \cite{BD}. Since $\partial_t$ is a Killing
field, $\langle T_{\mu t}\rangle$ is a conserved quantity. The
energy density seen by an observer with unit four-velocity
$u^{\mu}$ is then $u^{\mu}\langle T_{\mu t}\rangle/c^2$ (we
express $t$ in units of time). For simplicity, we will analyze
only black holes with spherical symmetry. In this case, the only
non-vanishing components of $\langle T_{\mu t}\rangle$ in the
asymptotic region (adopting spherical coordinates for $O$) are
$\langle T_{tt}\rangle=-c\langle T_{rt}\rangle=\pi c T^2/(12\hbar
l_P^2)$ \cite{BD}. The energy density for $O$ is determined by
$\rho=\langle T_{tt}\rangle/c^2$. On the other hand, since the
four-velocity of $O^{\prime}$ is $(\gamma,\gamma\vec{V})$ [in
$(t,\vec{x})$ coordinates], the density detected by the boosted
observer is precisely $\rho^{\prime}=\alpha \rho$.

Suppose now that, given the fundamental constants $\hbar$ and $c$,
the observer $O^{\prime}$ used the relation $l_P^2=\pi
T^2/(12\rho\hbar c)$ to determine the Planck length, but taking as
experimental values of the energy density and the temperature its
records $\rho^{\prime}$ and $T^{\prime}$. She will arrive at the
apparent Planck length $l_P^{\prime}=\sqrt{\alpha}l_P$. On the
other hand, if the black hole entropy has a statistical mechanical
origin, its value must be given by the logarithm of the number of
allowed quantum states. This number of degrees of freedom must not
depend on the Lorentz frame considered, in the same way as we have
shown that it is independent of the conformal frame. Then, $O$ and
$O^{\prime}$ must agree in the value of the entropy,
$S^{\prime}=S$. From this perspective, by using the BH formula the
boosted observer would assign the effective area
$A^{\prime}=\alpha A$ to the black hole horizon.

It is worth remarking that $A^{\prime}$ is only an effective,
operationally defined area, that differs from $A$ as a result of
the redshift caused by the boost. Actually, the true area of the
black hole horizon in the boosted frame continues to be $A$, but
this area is not directly observable from the asymptotic flat
region. To see that the geometric area of the horizon does not
change, let us analyze the most general stationary vacuum black
hole, namely the Kerr solution, expressed in Boyer-Lindquist
coordinates $(t,r,\theta,\phi)$ \cite{Wald}, where $t$ coincides
with the time employed in the asymptotic region by $O$. The
horizon has the topology of $S^2\times \mathbb{R}$. The sphere
$S^2$ admits the coordinates $\theta$ and $\psi=\phi-\Omega_H t\in
S^1$, while $\mathbb{R}$ is a null direction that can be described
by the coordinate $t$. In this coordinate system, the degenerate
metric on the horizon is diagonal and $t$-independent. Here,
$\Omega_H$ is the constant coordinate angular velocity on the
horizon \cite{Wald}. The areas that we want to compare are those
of the horizon sections of constant $t$ and $t^{\prime}$ (for $O$
and $O^{\prime}$). In principle, one might think of a possible
discrepancy between these areas because, on the sections of
constant $t^{\prime}$, the coordinate $t$ might vary modifying the
area element. However, this does not occur, since $t$ corresponds
to a null direction. Hence, the geometric area remains indeed the
same.

Summarizing, a boosted observer in the asymptotic flat region that
could carry out only local measurements on the Hawking radiation
would infer an effective value of the Planck length that is either
contracted or dilated. The value of the black hole entropy,
nonetheless, should be invariant. In addition, we emphasize that
the boost does not affect the geometric area of the horizon, in
contradistinction to the Fitzgerald-Lorentz contraction
experimented by the area of surfaces in flat spacetime \cite{SR}.

Finally, we consider changes of the Immirzi parameter in Ashtekar
gravity. In terms of the spacetime metric, the change from the
unit parameter to $\beta$ can be realized as the transformation
from the element $ds^2$ (with induced three-metric $h_{ij}$, lapse
$N$, and shift vector $N^{i}$) to \cite{LG}
\begin{equation}\label{immi}
d\bar{s}^2=-\beta^{-2}N^2dt^2+h_{ij}(dx^i+N^idt)(dx^j+N^jdt).
\end{equation}
Since the induced metric is not modified, the transformation
preserves the area of all constant-$t$ surfaces in the classical
theory and, in particular, the geometric area $A$ of a black hole
horizon. The scaling of the lapse has nevertheless an important
consequence, namely, it changes the normalization of any
(asymptotic) timelike Killing field. As a result, energies are
affected by a redshift according to the Tolman relation
$\bar{E}=\sqrt{(g_{tt}/\bar{g}_{tt})}E$, where $g_{\mu\nu}$
denotes the spacetime metric. In the case of Hawking radiation,
the shifted temperature and energy density are then $\bar{T}=\beta
T$ and $\bar{\rho}=\beta \rho$. Here, we have taken as reference
values those of the original frame with unit Immirzi parameter. We
therefore find a complete parallelism with the situation described
for boosts if we identify the parameters $\alpha$ and $\beta$.
Note, in addition, that both parameters are positive but otherwise
arbitrary.

In loop quantum gravity, as we have commented, the black hole
entropy can be calculated counting quantum states, obtaining
$S=\beta_0A_{\beta}/(4\beta l_{\star}^2)$. Although the horizon
area $A_{\beta}$ is classically independent of the Immirzi
parameter, its discrete quantum spectrum is proportional to this
constant. From this viewpoint, the area eigenvalues transform
under a change of $\beta$ exactly as the value of the effective
area $A^{\prime}$ under a change in the parameter $\alpha$
associated with Lorentz boosts. Furthermore, from the expression
of the black hole entropy and the BH formula, each Immirzi
description can be regarded as leading to a different effective
value $l_P(\beta)=l_{\star}\sqrt{\beta/\beta_0}$ for the Planck
length \cite{LG}. Introducing the definition $l_P\equiv
l_{\star}/\sqrt{\beta_0}$ as a fixed reference value, we then get
the transformation rule $l_P(\beta)=\sqrt{\beta}l_P$. In this way,
the scaling laws for our effective quantities reproduce those
deduced for boosts with the identification of parameters $\alpha$
and $\beta$.

This parallelism suggests the possibility that distinct Immirzi
formulations correspond in fact to descriptions associated with
different boosted frames. In this sense note that, as remarked in
Ref. \cite{AC}, a Lorentz boost connects different canonical
theories, just like a change of Immirzi description. According to
our conjecture, each Immirzi representation in loop quantum
gravity would provide an area spectrum measured in terms of a
different effective Planck length. The definition of this
effective length incorporates the gravitational redshift
introduced by the Immirzi parameter. This redshift is the same
that would affect the local measurements of a boosted observer in
a locally flat region where Lorentz invariance can be regained. In
this sense, the quantity that should be Lorentz invariant is not
the apparent value of the area (for any surface determined in a
dynamically consistent way), but rather the corresponding number
of elementary ``geometric excitations'', which is a dimensionless
variable. This is exactly what happens with the Immirzi ambiguity,
since the area spectrum differs only in the apparent value of the
dimensionful quantum of area. It is precisely because of this
property that the black hole entropy, when expressed as a
dimensionless quantity, becomes independent of the selected
Immirzi description.

Composing the above transformation laws, we can check that the
introduction of the Immirzi parameter $\beta$ followed by a
conformal scaling with factor $\Omega=\beta^{-1/2}$ indeed leaves
the kinematics of Ashtekar gravity invariant \cite{LG}. The
horizon area (spectrum) and the Planck length, which are
quantities that can be defined in a kinematical context \cite{LG},
remain unaltered by the composed transformation. The Hawking
temperature and energy density are however modified, since they
involve a concept of energy. This result and the analogy between
boosts and changes of Immirzi description imply also that there
must exist a composition of a boost and a conformal dilatation
that preserves the value of the Planck scale, in agreement with
recent works \cite{MS}.

In conclusion, we have argued that the value of the Planck length
determined by local measurements in a vacuum black hole spacetime
depends on the conformal and Lorentz frame of the asymptotic
observer. The value of the black hole entropy, on the other hand,
must be frame-independent, because it only accounts for the number
of quantum degrees of freedom compatible with the macroscopic
configuration. Via the BH relation, the effective ambiguity in the
Planck length then translates into an apparent ambiguity in the
dimensionful magnitude of the horizon area. This ambiguity is
similar to that caused by the Immirzi parameter in loop quantum
gravity. We have seen that the kinematical effects produced in
black hole physics by a boost, a constant conformal scaling, and a
change of the Immirzi parameter are in fact analogous. For the
case of the scaling, the analogy is broken as soon as dynamical
considerations (such as measurements of energy) are brought into
the scene. Remarkably, a complete parallelism is however
encountered between the consequences of Lorentz boosts and
transformations between Immirzi formulations. Based on this
result, we have suggested the possibility that the various Immirzi
representations of loop quantum gravity provide physical
descriptions in different boosted frames, each of them associated
with a different effective Planck length that takes into account
the redshift introduced. We believe that this conjecture and its
implications for Lorentz invariance and quantum gravity deserve
further examination which will be carried out elsewhere.

Finally, note that our arguments have been restricted to vacuum
black hole spacetimes, which possess an asymptotic flat region
where invariance under boosts is attained. Such an asymptotic
region needs not exist in more general spacetimes with a Killing
horizon, e.g. when the spatial hypersurfaces are compact. In this
latter case, however, the definitions of energies and temperatures
are already affected by an ambiguity in a multiplicative constant
at an early stage of the discussion, because there is no natural
way to fix the normalization of any timelike Killing vector in the
absence of the asymptotic region. Actually, this freedom in the
normalization translates into the possibility of scaling the null
normal to the horizon by a constant, a scaling that can be
interpreted as the result of a boost, but now performed at the
horizon.

The authors acknowledge DGESIC for financial support under
Research Project No. BFM2001-0213.


\small
\begin{thebibliography}{30}
\bibitem{Wald} Wald R M 1984 {\it General Relativity} (Chicago:
University of Chicago Press)

\bibitem{BCH} Bardeen J M, Carter B and Hawking S W 1973 {\it
Commun. Math. Phys.} {\bf 31} 161

\bibitem{Tbec} Bekenstein J D 1973 {\it Phys Rev.} D {\bf 7} 2333

Bekenstein J D 1974 {\it Phys Rev.} D {\bf 9} 3292

\bibitem{Thaw} Hawking S W 1975 {\it Commun. Math. Phys.} {\bf 43}
199

\bibitem{BraCo} Strominger A and Vafa C 1996 {\it Phys. Lett.} B
{\bf 379} 99

Maldacena J and Strominger A 1996 {\it Phys. Rev. Lett.} {\bf 77}
428

\bibitem{Eash} Ashtekar A, Baez J, Corichi A and Krasnov K 1998
{\it Phys. Rev. Lett.} {\bf 80} 904

Ashtekar A, Baez J C and Krasnov K 2001 {\it Adv. Theor. Math.
Phys.} {\bf 4} 1

\bibitem{Ash} Ashtekar A 1991 {\it Lectures on Non-Perturbative
Canonical Gravity} ed L Z Fang and R Ruffini (Singapore: World
Scientific)

Thiemann T 2001 Introduction to Modern Canonical Quantum General
Relativity {\it Preprint} gr-qc/0110034

\bibitem{Immi} Immirzi G 1997 {\it Nucl. Phys. Proc. Suppl.} {\bf 57}
65

Immirzi G 1997 {\it Class. Quantum Grav.} {\bf 14} L177

\bibitem{SN} Rovelli C and Smolin L 1995 {\it Phys. Rev.} D
{\bf 52} 5743

Rovelli C 1998 {\it Living Rev. Rel.} {\bf 1} 1

\bibitem{RS} Rovelli C and Smolin L 1995 {\it Nucl. Phys.} B
{\bf 442} 593

Rovelli C and Smolin L 1995 {\it Nucl. Phys.} B {\bf 456} 753

\bibitem{Aspec} Ashtekar A and Lewandowski J 1997 {\it Class.
Quantum Grav.} {\bf 14} A55

\bibitem{Cash} Ashtekar A, Corichi A and Krasnov K 2000 {\it Adv.
Theor. Math. Phys.} {\bf 3} 419

\bibitem{Imvar} Rovelli C and Thiemann T 1998 {\it Phys. Rev.}
D {\bf 57} 1009

Gambini R, Obreg\'{o}n O and Pullin J 1999 {\it Phys. Rev.} D {\bf 59}
047505

Samuel J 2001 {\it Phys. Rev.} D {\bf 64} 048501 (2001)

Mena Marug\'{a}n G A 2002 {\it Class. Quantum Grav.} {\bf 19} L63

\bibitem{LG} Garay L J and Mena Marug\'{a}n G A 2002 {\it Phys. Rev.}
D {\bf 66} 024021

\bibitem{Alex} Alexandrov S 2000 {\it Class. Quantum Grav.} {\bf 17}
4255

Alexandrov S 2002 {\it Phys. Rev.} D {\bf 65} 024011

\bibitem{BD} Birrell N D and Davies P C W 1986 {\it Quantum
Fields in Curved Space} ed P V Landshoff {\it et al.} (Cambridge:
Cambridge University Press)

\bibitem{dcon} Duff M J, Okun L B and Veneziano G 2002 {\it JHEP}
{\bf 0203} 023

\bibitem{SR} French A P 1968 {\it Special Relativity} (New York:
Norton)

\bibitem{Note} To be more precise, we disregard measurements
that require an unbounded amount of time when the asymptotic flat
region is approached

\bibitem{AC} Amelino-Camelia G 2002 On the Fate of Lorentz Symmetry in
Loop Quantum Gravity and Noncommutative Spacetimes {\it Preprint}
gr-qc/0205125

\bibitem{MS} Magueijo J and Smolin L 2002 {\it Phys. Rev. Lett.} {\bf 88}
190403
\end{thebibliography}
\end{document}